\documentclass[fleqn,10pt,a4paper]{article}
\usepackage{amsmath,amssymb,bm}
\usepackage{cite,epsfig,color,graphicx}

\begin{document}

\title{Non-linear electrostatic waves in Born-Infeld plasmas}


\author{DA Burton\thanks{Department of Physics, Lancaster University, UK 
and the Cockcroft Institute, Daresbury, UK}
\and 
H Wen\footnotemark[1]
}

\maketitle

\begin{abstract}
Motivated by the suggestion that Born-Infeld plasmas could have significance for
electron acceleration in neutron star crusts, we obtain an upper bound on
the amplitude of electrostatic waves propagating parallel to a longitudinal
magnetic field in a Born-Infeld plasma.
\end{abstract}

\section{Introduction}
Born-Infeld electrodynamics has attracted
considerable interest over recent years. It was originally introduced in the
1930s~\cite{born:1934} as an attempt to describe the classical electron entirely
in terms of its electromagnetic field, but interest in it soon waned in favour
of quantum theory. However, during the first superstring revolution in the
mid-1980s it was shown that Born-Infeld-type theories are a feature of low
energy string field theory~\cite{fradkin:1985}, and this discovery led to the
resurgence of interest in Born-Infeld electrodynamics seen
recently~\cite{gibbons:2001, ferraro:2007,
ferraro:2010, dereli:2010, burton:2010}.

Furthermore, among the family of non-linear
generalizations of Maxwell electrodynamics, it has long been known
that Born-Infeld theory possesses a number of highly attractive features; in
particular, like the vacuum Maxwell equations, the Born-Infeld equations exhibit
zero birefringence and its solutions have exceptional causal
behaviour~\cite{boillat:1970, plebanski:1968}.
The vacuum Maxwell and Born-Infeld field equations are the only
physical theories with a single light cone obtainable from a local Lagrangian
constructed solely from the two Lorentz invariants associated with the
electromagnetic field strength tensor and the metric tensor.

Any self-consistent theory describing a large collection of charged particles
must include all electromagnetic forces between the particles. However, the
notorious problem of determining the classical force on a single accelerating
point charge due to its own electromagnetic field has stimulated research for
over a century and remains unresolved. The structure of an isolated single
electron is currently beyond observation and one often proceeds classically by
associating the electron with a singularity in the electromagnetic field
described by Maxwell's equations in vacuo. Following Dirac~\cite{dirac:1938}, an equation of
motion for the electron may be obtained by appealing to conservation of
the total energy-momentum of the electron and its electromagnetic field
(see~\cite{burton:2007} for a recent discussion). In
order to remove singularities in the equation of motion, Dirac made ``natural
assumptions'' about the origin of the electron mass. The resulting Lorentz-Dirac
equation of motion contains \emph{third} order proper time derivatives of the
electron's world line, and possesses solutions that violate
intuition. In particular, unless
special conditions are adopted for the final state of the electron, it predicts
that a free electron in vacuo can self-accelerate; furthermore the equations
possess solutions in which the electron experiences a sudden acceleration
before it enters a region of space containing a non-vanishing external
electrostatic field (see~\cite{griffiths:2010} for a recent discussion).

The search for a complete dynamical theory of a point charge within
Born-Infeld electrodynamics is on-going~\cite{cruscinski:1998}, and
this theory is expected to provide a resolution to the radiation-reaction
problem. In particular, the difficulties associated with the Lorentz-Dirac
equation are thought to have their origin in the electron's singular total
mass-energy in classical Maxwell electrodynamics; however, the electric field of
a Born-Infeld electron at rest is non-singular and its total mass-energy is
finite.

Some of the most extreme conditions ever encountered in a terrestrial
laboratory are created when high-power laser pulses interact with
matter. The laser pulse immediately vaporizes the matter to form an
intense laser-plasma providing novel avenues for generating intense
bursts of coherent electromagnetic radiation for a wide range of
applications in biological and material science~\cite{schlenvoigt:2008}.
Furthermore, laser-plasmas permit controllable investigation of matter in extreme
conditions that only occur naturally away from the Earth.
It is expected that the next generation of ultra-intense
lasers will, for the first time, allow controllable access to regimes
where a host of different quantum electrodynamic phenomena will be
evident. In particular, the challenge of extending Schwinger's classic
analysis~\cite{schwinger:2007} of vacuum breakdown in a static external electric
field to breakdown in an intense laser-plasma is on-going~\cite{marklund:2009}.
However, the radiation-reaction problem is sufficiently strong motivation for
exploring whether a Born-Infeld-type theory can yield experimental signatures
{\it before} quantum effects become significant~\cite{dereli:2010}.
A sufficiently short and intense laser pulse propagating through a
plasma may create a travelling longitudinal plasma wave whose phase velocity
is approximately the same as the laser pulse's group
velocity. However, it is not possible to sustain arbitrarily large electric
fields; substantial numbers of plasma electrons become trapped in the wave and
are accelerated, which dampens the wave (the wave `breaks'). Early theoretical
investigation of non-linear plasma waves was undertaken in the mid 1950s
by Akhiezer and Polovin~\cite{akhiezer:1956}, and later expounded by
Dawson~\cite{dawson:1959} in the context of wave-breaking. Furthermore, this
acceleration mechanism was recently employed~\cite{diver:2010} to explain the
emission of energetic electrons from within the interiors of pulsars; such
electrons are necessary for the formation of the electron-positron plasma
populating a pulsar's magnetosphere. 

Wave-breaking is a fundamentally non-linear phenomenon, and it is
natural to explore the properties of Born-Infeld plasmas from this
perspective. Moreover, the magnetic fields found in neutron stars are
typically $\sim 10^8 {\rm T}$, whilst those in magnetars may be two orders of
magnitude higher and such fields have energy densities commensurate with the
Schwinger limit (i.e. commensurate with a static electric field of strength
$\sim 10^{18}\,{\rm V}/{\rm m}$). 

The following is a brief summary of our recent analysis of Born-Infeld
plasma waves near breaking. The present work is an application of the
approach established in~\cite{burton:2010}, generalized here to include
the presence of a background magnetic field. The notation and conventions used
here are identical to those in~\cite{burton:2010}, to which we refer the
reader for further details. In particular, we use units in which the speed of
light $c=1$ and the permittivity of the vacuum $\varepsilon_0=1$. 
\section{Born-Infeld plasma}
Let $({\cal M},g)$ be a spacetime. The electromagnetic sector of a Born-Infeld
plasma may be expressed in terms of the Lagrangian ${\cal L}_{\rm BI}(X,Y)$,
\begin{equation}
\label{definition_LBI}
{\cal L}_{\rm BI}(X,Y) = \frac{1}{\kappa^2}(1-\sqrt{1-\kappa^2 X -
  \kappa^4 Y^2/4}),
\end{equation}
where the invariants $X$ and $Y$ are
\begin{equation}
\label{invariants}
X = \star (F\wedge \star F),\qquad Y = \star (F\wedge F),
\end{equation}
with $\star$ the Hodge map associated with the spacetime metric $g$ and $\kappa$
is a constant that characterizes the self-interaction of the electromagnetic
field. The relationship between the excitation $2$-form $G$ and the
Maxwell $2$-form $F$ is
\begin{align}
\notag
G &= 2 \bigg(\frac{\partial{\cal
      L}_{\rm BI}}{\partial X} F -\frac{\partial{\cal
      L}_{\rm BI}}{\partial Y} \star F\bigg)\\
\label{def_G}
  &= \frac{1}{\sqrt{1-\kappa^2 X - \kappa^4 Y^2/4}}\bigg(F -
\frac{\kappa^2 Y}{2}\star F\bigg)
\end{align}
and Maxwell electrodynamics (where $G=F$) is recovered in the limit $\kappa
\rightarrow 0$.

For simplicity, the plasma electrons are represented as a {\it cold} relativistic
fluid; their worldlines are trajectories of a unit normalized future-pointing
timelike $4$-vector field $V$ satisfying 
\begin{equation}
\label{lorentz}
\nabla_V \widetilde{V} = \frac{q}{m}\iota_V F,\qquad g(V,V)=-1
\end{equation}
where $q<0$ is the charge on the electron, $m$ is
the electron rest mass, $q \iota_V F$ is the Lorentz $4$-force acting on the electron fluid, $\iota_V$ is the interior product with respect to $V$ and $\nabla$ is the Levi-Civita
connection on ${\cal M}$. The $1$-form $\widetilde{V}$ is the metric dual of
the vector field $V$, i.e. the $1$-form $\widetilde{V}$ satisfies
$\widetilde{V}(U)=g(V,U)$ for all vector fields $U$ on ${\cal M}$.

We are interested in the evolution of a plasma over timescales during
which the motion of the ions is negligible in comparison with the
motion of the electrons. Here the ions are specified as a background and their
worldlines are trajectories of the prescribed future-pointing timelike
$4$-vector field $N_{\rm ion}$ (the ion number $4$-current) on ${\cal M}$.

Maxwell's equations are
\begin{equation}
\label{maxwell}
dF = 0,\qquad d\star G = - q n\star\widetilde{V} - q_{\rm ion} \star\widetilde{N_{\rm ion}}
\end{equation}
where the $0$-form $n$ is the electron proper number density and $q_{\rm ion} =
Z|q|$ is the charge on an ion, with $Z$ the multiplicity of the ionization. The
$1$-form $\widetilde{N_{\rm ion}}$ is the metric dual of the vector field
$N_{\rm ion}$.
\section{Non-linear electrostatic waves}
Particle acceleration in non-linear electrostatic waves close to breaking has recently
been proposed as a possible mechanism for explaining how energetic electrons are
ejected from within the interiors of pulsars~\cite{diver:2010}.

If an atom is immersed in a uniform background magnetic field whose strength is
much greater than $\sim 10^5 {\rm T}$ then the corresponding magnetic force on
the electrons is much greater than the atom's Coulombic forces
\cite{harding:2006}. The atom settles into the ground
Landau level, limiting the electrons' spatial displacement
transverse to the magnetic field lines. Thus, electrons are conducted
preferentially along the direction of the magnetic field lines, and one may
approximate the bulk electron motion as 1-dimensional~\cite{diver:2010}.
Moreover, the magnetic field lines in the iron crust of a neutron star are
expected to run parallel to its surface and to be strongly curved near the
poles, where they emerge normal to the star's surface. The
magnetic curvature near the poles is expected to lead to variations
in electron number density and excite electrostatic waves in
the electron `gas' within the iron crust~\cite{diver:2010}.   

To obtain some insight into the behaviour of a Born-Infeld plasma in this
context, in the following we explore properties of solutions to (\ref{lorentz}),
(\ref{maxwell}) that describe large-amplitude longitudinal electrostatic waves
propagating in a uniform background ion density in a flat spacetime. It is
useful to envisage the electrons in the plasma as belonging to one or other of
two families. The first family and the ion background are represented by the
triple $\{V,n,N_{\rm ion}\}$ and constitute the bulk plasma; those electrons and
the background ions form the electrostatic wave. The members of the second
family are the rest of the electron population, some of which are trapped in the
wave's potential; we do not attempt to include the second family in the simple
model explored here. 

Let $(x^a)$ be an inertial coordinate system on Minkowski spacetime
$({\cal M},g)$ where $x^0$ is the proper time of observers at fixed
Cartesian coordinates $(x^1,x^2,x^3)$. The metric tensor
$g$ has the form
\begin{equation}
\label{metric}
g = \eta_{ab}\, dx^a \otimes dx^b
\end{equation}
with
\begin{equation}
\label{eta_ab}
\eta_{ab} =
\begin{cases}
-1\,\, {\rm if}\,\, a=b=0,\\
1\,\, {\rm if}\,\, a=b\neq 0,\\
0\,\, {\rm if}\,\, a \neq b
\end{cases}
\end{equation}
and the Hodge map $\star$ is induced from the $4$-form $\star 1$ on ${\cal
  M}$ where
\begin{equation}
\label{volume_4-form}
\star 1 = dx^0 \wedge dx^1 \wedge dx^2 \wedge dx^3.
\end{equation}
The ion background is given as the number $4$-current $N_{\rm ion} = n_{\rm
ion} \partial/\partial x^0$, where $n_{\rm ion}$ is a constant.

The longitudinal electrostatic waves considered here propagate parallel to the
$x^3$-axis with phase velocity $v$  (with $0 < v <1$) in the ion rest frame.
We introduce the pair $\{e^1,e^2\}$
\begin{equation}
\label{longitudinal_co-frame}
e^1= vdx^3 - dx^0, \qquad e^2=dx^3 - vdx^0
\end{equation}
and note that the orthonormal co-frame $\{\gamma e^1,\gamma e^2,
dx^1, dx^2\}$ is adapted to observers moving at velocity $v$ along
$x^3$ (observers in the `wave frame'), where the Lorentz factor
$\gamma=1/\sqrt{1-v^2}$.

We seek a unit normalized $4$-velocity field $V$ of the form
\begin{equation}
\label{V_ansatz}
\widetilde{V} = \mu(\zeta)\, e^1 -\sqrt{\mu(\zeta)^2-\gamma^2}\, e^2
\end{equation}
where $\zeta = x^3 - vx^0$ is the wave's phase and $e^2 = d\zeta$. We
have adopted the so-called `quasi-static approximation'; the
pointwise dependence of $\mu$ is on $\zeta$ only. The velocity of the bulk
plasma electrons observed in the wave frame is $\sqrt{\mu^2-\gamma^2}/\mu$
in the direction of {\it decreasing} $x^3$, so the electrons move slower
than the wave propagates. The maximum amplitude wave arises for oscillations
during which $\mu$ is arbitrarily close to $\gamma$ (i.e. the bulk plasma
electrons catch the wave).

The Maxwell $2$-form $F$ is
\begin{equation}
\label{Maxwell_2-form}
F = E(\zeta)\, dx^0 \wedge dx^3 - B\, dx^1 \wedge dx^2  
\end{equation}
where $E$ is the $x^3$ component of the electric field and its
pointwise dependence is on $\zeta$ only. The constant $B$ is the component of
the magnetic field along $x^3$, and the remaining components of the electric
and magnetic field vanish.

Employment of (\ref{metric})-(\ref{Maxwell_2-form})
reduces (\ref{def_G})-(\ref{maxwell}) to a second-order ordinary differential
equation for $\mu$, viz.
\begin{equation}
\label{ODE_for_mu}
\frac{d}{d\zeta}
\bigg(\frac{E}{\sqrt{1-\kappa^2 E^2}}\bigg)
= \frac{q\, Z\, n_{\rm ion}\, \gamma^2}{\sqrt{1+\kappa^2 B^2}}
\bigg(\frac{v\,\mu}{\sqrt{\mu^2-\gamma^2}} - 1\bigg),
\end{equation}
where
\begin{equation}
E = \frac{1}{\gamma^2}\frac{m}{q}\frac{d\mu}{d\zeta}.
\end{equation}

\begin{figure}
\begin{center}
\includegraphics{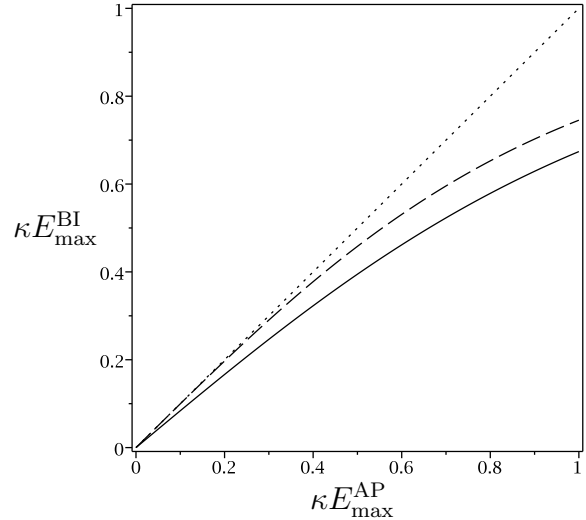} 
\caption{\label{fig:BI_Emax} A graph of the maximum amplitude $E^{{\rm BI}}_{\rm
max}$ of an electrostatic wave in a Born-Infeld plasma versus the maximum
amplitude $E^{{\rm AP}}_{\rm max}$ of a Maxwell plasma. Both field strengths
are scaled by the Born-Infeld constant $\kappa$. The dashed curve corresponds to
$B=0$ and the solid curve corresponds to $B = 1/(\kappa c)$. The dotted line
lies along the diagonal.}
\end{center}
\end{figure}

Inspection of (\ref{ODE_for_mu}) reveals that, for oscillatory solutions, the
electric field $E$ has a turning point where $\mu=\gamma^2$. Further
investigation reveals that
this turning point is a minimum of $E$ (for $q<0$), and an upper bound $E_{\rm
max}$ (`wave-breaking limit') on $E$ may be obtained by evaluating the first
integral of (\ref{ODE_for_mu}) between the points $(\mu=\gamma,\, E=0)$ and
$(\mu=\gamma^2,\, E=-E^{\rm BI}_{\rm max})$. Restoring the speed of light $c$
and permittivity of the vacuum $\varepsilon_0$ yields
\begin{equation}
\label{AP_limit}
E^{\rm BI}_{\rm max} = \frac{1}{\kappa}
\sqrt{1 - \bigg[
\frac{\kappa^2\, E^{{\rm AP}\,\,2}_{\rm max}}
{2\sqrt{1+\kappa^2\, c^2\, B^2}}
+ 1\bigg]^{-2}}
\end{equation}
where $E^{\rm AP}_{\rm max}$ is the maximum electric field for
a relativistic cold Maxwell plasma,
\begin{equation}
E^{\rm AP}_{\rm max} = \frac{m\omega_p c}{|q|}\sqrt{2(\gamma - 1)}
\end{equation}
first obtained by Akhiezer and Polovin~\cite{akhiezer:1956}, and the constant
$\omega_p$ is the plasma frequency
\begin{equation}
 \omega_p = \sqrt{\frac{q^2 Z n_{\rm ion}}{m \varepsilon_0}}.
\end{equation}
If $\kappa = 10^{-18} {\rm m}/{\rm V}$ then $\kappa c
B = 1$ corresponds to $B = 10^{9} {\rm T}$, which is
within the range of the surface magnetic fields of rotation-powered
radio pulsars~\cite{harding:2006}. The relationship between $\kappa
E^{\rm BI}_{\rm max}$ and $\kappa E^{{\rm AP}}_{\rm max}$ is shown in
figure~\ref{fig:BI_Emax} for $B=0$ and $B = 1/(\kappa c)$.
%

\section{Acknowledgments}
We thank the Cockcroft Institute and the ALPHA-X project for support.

\end{document}